**Computational Study of Tunneling Transistor Based on Graphene Nanoribbon**

Pei Zhao, Jyotsna Chauhan, and Jing Guo*
Department of Electrical and Computer Engineering
University of Florida, Gainesville, FL, 32611-6130

ABSTRACT

Tunneling field-effect transistors (FETs) have been intensely explored recently due to its potential to address power concerns in nanoelectronics. The recently discovered graphene nanoribbon (GNR) is ideal for tunneling FETs due to its symmetric bandstructure, light effective mass, and monolayer-thin body. In this work, we examine the device physics of p-i-n GNR tunneling FETs using atomistic quantum transport simulations. The important role of the edge bond relaxation in the device characteristics is identified. The device, however, has ambipolar *I-V* characteristics, which are not preferred for digital electronics applications. We suggest that using either an asymmetric source-drain doping or a properly designed gate underlap can effectively suppress the ambipolar *I-V*. A subthreshold slope of 14mV/dec and a significantly improved on-off ratio can be obtained by the p-i-n GNR tunneling FETs.

* guoj@ufl.edu

With the scaling down of field-effect transistors (FETs), power dissipation has been increasing dramatically.[1] At the same time, the static power dissipation is becoming an increasingly important concern. For lowering the static power dissipation, the most important parameter for optimization is the subthreshold swing (SS), which is the voltage required to change current by an order of magnitude. In a metal-oxide-semiconductor (MOS) FETs, the value has been limited by thermionic emission to 60mv/dec at room temperature,[2] and it is getting worse as transistors are scaling down in size. In recent years, there has been a persistent pursuit for alternative device structures and materials that could provide a subthreshold swing less than 60mv/dec. A large number of transistors have been reported that explore band-to-band tunneling principles in silicon, germanium, and carbon nanotube (CNT) to obtain the subthreshold swing less than 60 mv/dec and high $I_{on}/I_{off}$ ratio by experiment and theory.[3-10] Studies on the GNR MOSFETs using both tight-binding (TB) and first-principles methods have been reported.[11-14] This letter presents a computational study of the p-i-n tunneling FETs using a graphene nanoribbon (GNR) as the channel material, which has rarely been studied before. The recently discovered GNR is a preferred material of choice due to its symmetrical band structure, light effective mass, and direct band gap to favor tunneling.[15-18] Fundamental questions regarding the important role of GNR edges, schemes for controlling ambipolar transport, and achievable device performance, however, remain unclear.

In this letter, we theoretically explore the device physics of GNR tunneling FETs by using three-dimensional atomistic simulations. We show that the edge bond relaxation has a significant effect on the device characteristics of the p-i-n GNR tunneling FETs, which distinguishes it from a CNT tunneling FETs. A subthreshold swing of 14mV/dec and a large on-off ratio are obtained at the ballistic transport limit in the presence of edge bond relaxation. The device, however, shows ambipolar *I-V* characteristics that are not preferred for digital electronics applications. We show that by using an asymmetric source-drain doping or a properly designed gate underlap, the ambipolar characteristics can be significantly suppressed. The modeled p-i-n GNR tunneling FETs has a double-gate geometry with a gate oxide thickness of 1.5 nm and the dielectric constant of κ= 16 (for $HfO_2$), as schematically shown in Figure 1. A semiconducting armchair-edge GNR (AGNR) is used as the channel material.[19-21] The AGNR has an index of $n$=13, which results in a width of ~1.6nm and a bandgap of ~0.86 eV. The AGNR channel is intrinsic, and has the same length as the gate, $L_{ch}$=30nm. The p-type doping density of the semi-infinite source extension is 0.01 dopant/atom, and the drain extension is n-type doped to the same density. The workfunction of the gate electrode is adjusted to a value so that the minimal leakage current appears at $V_G$=0, and a variation of the gate workfunction results in a shift of the threshold voltage. A power supply voltage of $V_{DD}$ = 0.4 V and room temperature operation $T$=300K are assumed. The above parameters are nominal ones, and we explore various issues by varying the parameters.

To model the device characteristics, open-boundary Schrodinger equation is solved in an atomistic $p_Z$ orbital basis set using the non-equilibrium Green's function (NEGF) formalism.[22] A nearest neighbor tight binding (TB) parameter of $t_0$ = -2.7 eV is used. Previous *ab initio* simulations, however, indicated the important role of the edge bond relaxation in the AGNR, which changes the edge bond length and bond parameters.[23-24] To model edge bond relaxation, we use a different TB parameter, $t_0^{'} = C_{edge} t_0$ for the edge bonds, where $C_{edge}$=1.12 as parameterized to the *ab initio* bandstructure simulations.[23] Ballistic transport is assumed. As indicated by a study on the p-i-n CNT tunneling FETs, phonon scattering has a small effect on the p-i-n tunneling FETs characteristics if the hot phonon effect is small.[9] The semi-infinite source and drain contacts are accounted for by the contact self-energy matrix, which is solved by a recursive relation. The atomistic transport equation is self-consistently solved with a three-dimensional (3D) Poisson equation using the finite element method. The gate leakage current, which can be suppressed by increasing the gate insulator thickness, is neglected here. The noise due to thermal effects and nonideal trap states is not dealt with here for simplicity.

We first examine the effect of edge bond relaxation on device performance. Figure 2 compares the log($I_D$) *vs.* $V_G$ characteristics at $V_D$=0.4V in the presence and in the absence of edge bond relaxation. The 13-AGNR, as shown in Figure 2(a), is a representative case for the $n=3p+1$ GNR group, in which the edge bond relaxation results in an increase of the bandgap. In contrast, the 12-AGNR, as shown in Figure 2(b), is a representative case but for the n=$3p$ GNR group, in which the edge bond relaxation results in a decrease of the bandgap.[23] As shown in Figure 2(a), in the presence of edge bond relaxation, the minimal leakage current $I_{min}$ decreases by about $10^6$ and the on-current (defined at $V_G$=$V_D$=$V_{DD}$=0.4V) decreases by about a factor of 5. The subthreshold swing decreases from $S$=51mV/dec to $S$=14mV/dec. The results indicated the important role of edge bond relaxation, which must be considered in designing GNR tunneling FETs. The qualitative trend is opposite for the $n$=3$p$ GNR group. As shown in Figure 2(b), the minimal leakage current and the subthreshold swing increase considerably for the 12-AGNR in the presence of edge bond relaxation. In comparison, it is not an issue for a CNT tunneling FETs, which does not have an edge. The significant decrease (increase) of $I_{min}$ and $S$ is due to the increase (decrease) of the bandgap stemming from the edge bond relaxation. As shown in inset of Figure 2(a), at minimal leakage point ($V_G$=0V) without edge bond relaxation bandgap for 13-AGNR is 0.71eV, the bottom of conduction band edge in the channel region is lower than top of valence band edge in the source region. Band to band tunneling current exists even at the minimal bias condition. When considered edge bond relaxation, the bandgap increases to 0.86eV, the bottom of conduction band edge in the channel region becomes higher than top of valence band edge in source region. As a result, the band-to-band tunneling is completely turned off. The same argument applies to the drain-channel junction, which is symmetric to the source-channel junction at the minimal bias condition. Although the increase of the bandgap from 0.71eV to 0.86eV is slight, complete turn-off of the band-to-band tunneling results in a significant lowering of the minimal leakage current by about six orders of the magnitude. The subthreshold swing is smaller than 60mV/dec because the exponentially decaying tail of the electron distribution functions is cut off by the semiconductor band gap in the source and drain extensions. The device also shows ambipolar $I_D$-$V_G$ characteristics, due to a similar reason as the ambipolar characteristics in Schottky barrier GNRFETs or CNTFETs. Band-to-band tunneling from the source to the channel results in electron conduction at high gate voltages, and that from the drain to the channel results in hole conduction at low gate voltages. The device can be conceptually viewed as an n-type FET (due to source-channel tunneling) in parallel with a p-type FET (due to drain-channel tunneling). Ambipolar *I-V* characteristics, however, are not preferred for digital electronics applications.[25] We propose and examine two schemes for suppressing ambipolar characteristics in the p-i-n GNR tunneling FETs.

Controlling the ambipolar characteristics by designing the drain doping density is examined next. If the doping of the drain extension is reduced, the width of the band-to-band tunneling barrier for holes at the drain end of the channel increases due to a larger electrostatic screening length for a lower doping density, as shown in Figure 3(b). As a result, the p-type conduction branch is significantly suppressed, because tunneling current decreases exponentially with an increasing barrier width. In contrast, because the n-type conduction is controlled by the electron band-to-band tunneling from the source to the channel, it is insensitive to the drain doping density. Asymmetric source-drain doping, therefore, offers a successful scheme for suppressing ambipolar *I-V* characteristics, as shown in Figure 3(a). It is also observed that the source-drain current is nearly independent of the gate voltage in $-0.2V<V_G<0$ for a drain doping density of $N_D$=0.001 dopant/atom. The reason is that in this bias range, the band-to-band tunneling is completely turned off and the current is due to the direct source-drain tunneling, as indicated by the lack of the band-to-band tunneling peak in the current spectrum plot, compared to the case of higher drain doping densities shown in Figure 3(c).

We next examine controlling the ambipolar characteristics by using a gate underlap at the drain end of the channel, as shown in the inset of Figure 4(a). To examine the underlap effect, we fix the channel length at 50nm and vary the length of the gate underlap. Figure 4(a) shows a suppression of the ambipolar characteristics as the drain underlap increases. Again, the n-type conduction branch is unaffected and the p-type conduction branch is suppressed. The increase of the gate underlap at the drain end results in a nearly linear potential drop in the ungated part of the channel, which increases the width of the band-to-band tunneling barrier, as shown in Figure 4(b). The increase of the tunneling barrier thickness results in an exponential decrease of the band-to-band tunneling current, as shown by the current spectrum in Figure 4(c). We also emphasize that the gate length should not be shrunk so short that the direct source-drain tunneling becomes a concern. As shown in the inset of Figure 4(c), which zooms in the current spectrum due to direct source-drain tunneling, an increase of the drain underlap results in an increase of the direct source-drain tunneling current due to a thinner tunneling barrier attributed to a shorter gate length.  Figure 4 shows that a properly designed gate underlap can be another effective method to suppress ambipolar characteristics.

Finally we examine the device performance of the nominal GNR tunneling FETs as shown in Figure 1 at different supply voltage. To characterize the performance of the device, we use a previously developed scheme, which plots $I_{on}$ as a function of $I_{on}/I_{off}$ as shown in Figure 5.[26] A significant improvement in terms of the maximum achievable on-off ratio is observed, especially compared to Schottky barrier GNRFETs which suffers from small maximum achievable on-off ratio when the gate oxide thickness is scaled down. Because the thermionic emission tail in the source and drain regions are suppressed by the bandgap for the p-i-n GNR tunneling FETs, the minimal leakage current is small and the subthreshold swing can be considerably smaller than the 60mV/dec room temperature limit. As a result, when the power supply voltage is 0.3V, the maximum achievable $I_{on}/I_{off}$ ratio is up to $10^{11}$. The maximum on-off ratio decreases considerably as the power supply voltage increases above 0.4V, due to the turn on of the band-to-band tunneling at the minimal leakage bias condition, as compared to the direct source-drain tunneling as the only leakage mechanism for low $V_{DD}$. The maximum achievable $I_{on}/I_{off}$ of $10^6$ at $V_{DD}$=0.5V, however, is still significantly better than the value of 100 by a Schottky barrier GNRFETs with a similar channel material. The increase of the $I_{on}/I_{off}$, however, comes at an expense of a lower $I_{on}$ due to the existence of the band-to-band tunneling barrier at the source end in the on-state. Optimization of the on-current will require further engineering of the tunneling barrier at the source end, which is out of the scope of this letter.

The device characteristics investigated here is for an ideal smooth edge, which establishes the performance limits of the GNR p-i-n tunneling FETs. Although the edge quality still remains to be improved, recent experiments have made significant progress for achieving smooth-edge GNRs.[18,27] GNR edge roughness could affect the performance of the GNR tunneling FETs in two ways. First, bandgap states could be induced by the GNR edge roughness. The states, especially in the band-to-band tunneling junction regions, can assist tunneling and results in an increase of both the on-current and the leakage current. Second, the edge roughness can result in edge scattering for conducting electrons which lowers the current. A detailed study of the edge roughness effect in GNR tunneling FET is beyond the scope of this letter.

Demonstration of the GNR p-i-n tunneling FETs would require techniques for obtaining narrow GNRs and for doping GNRs developed. The recently demonstrated method for chemically deriving the GNRs from graphene is capable of producing GNRs down to a width of about 1.5nm.[18]Attaching functional groups to the chemically reactive edge of the GNRs could be a promising method for achieving the required doping in the GNR tunneling FETs.

In summary, device physics of GNR tunneling FETs is studied by 3D atomistic simulations. The important role of the edge bond relaxation on device characteristics is discovered. The modeled device shows a subthreshold swing of 14mV/dec at room temperature, and significantly improved on-off ratio, especially at low power supply voltages. The ambipolar *I-V* characteristics are a concern for digital electronics applications. We show that by either using an asymmetric source-drain doping or a properly designed gate underlap, the undesired ambipolar characteristics can be suppressed significantly for the p-i-n GNR tunneling FETs.

**Acknowledgement**

This work was supported in part by ONR and NSF. The simulations were performed in the University of Florida High Performance Computing Center.

## FIGURE CAPTIONS

Figure 1 *The modeled device structure.* The p-i-n GNR tunneling FETs has a double gate with the gate oxide thickness of $t_{ox}$=1.5nm and dielectric constant of $\kappa$=16. The nominal parameters are listed below. The $n$=13 armchair-edge GNR channel is intrinsic with a channel length of $L_{ch}$=30nm. The p-type doping density of the source extension is $N_A$=0.01 dopant/atom, and the n-type doping density of the drain is $N_D$=0.01 dopant/atom.

Figure 2 The log($I_D$) *vs.* $V_G$ characteristics of the GNR tunneling FETs as shown in Figure 1 at $V_D$=0.4V for (a) the $n$=13 AGNR channel and (b) the $n$=12 AGNR channel in the presence of edge bond relaxation (dashed lines) and in the absence of edge bond relaxation (solid lines). The band profiles at the source-channel junction are shown in the insets with the same symbols for the minimal bias condition ($V_G$=0).

Figure 3 *Effect of drain doping.* (a) The log($I_D$) *vs.* $V_G$ characteristics at $V_D$=0.4V for the GNR tunneling FETs as shown in Figure 1 with different drain doping densities (b) The band profiles and (c) the energy resolved current spectrum at $V_G$=-0.1V and $V_D$=0.4V. The source (drain) Fermi level, $E_{FS}$ ($E_{FD}$) is also shown in (b). The simulated drain doping densities are $N_D$=0.001 (dash-dot lines), 0.004 (dotted lines), 0.008 (dashed lines), and 0.01 (solid lines) dopant/atom. The source doping density is fixed at $N_A$=0.01 dopant/atom.

Figure 4 *Effect of gate underlap.* (a) The log($I_D$) *vs.* $V_G$ characteristics at $V_D$=0.4V for the GNR tunneling FETs as shown in Figure 1 with different gate underlap at the drain end of the channel as shown in the inset. The AGNR channel has an index of $n$=13 and a channel length of 50nm. (b) The band profiles and (c) the energy resolved current spectrum at $V_G=-0.4V$ and $V_D=0.4V$. The simulated gate underlap lengths are 0 (solid lines), 10nm (dashed lines), and 15nm (dotted lines).

Figure 5 *Device performance.* The on-current *vs.* on-off ratio for the GNR tunneling FETs as shown in Figure 1 in the presence of edge bond relaxation at different power supply voltages.

**FIGURES**

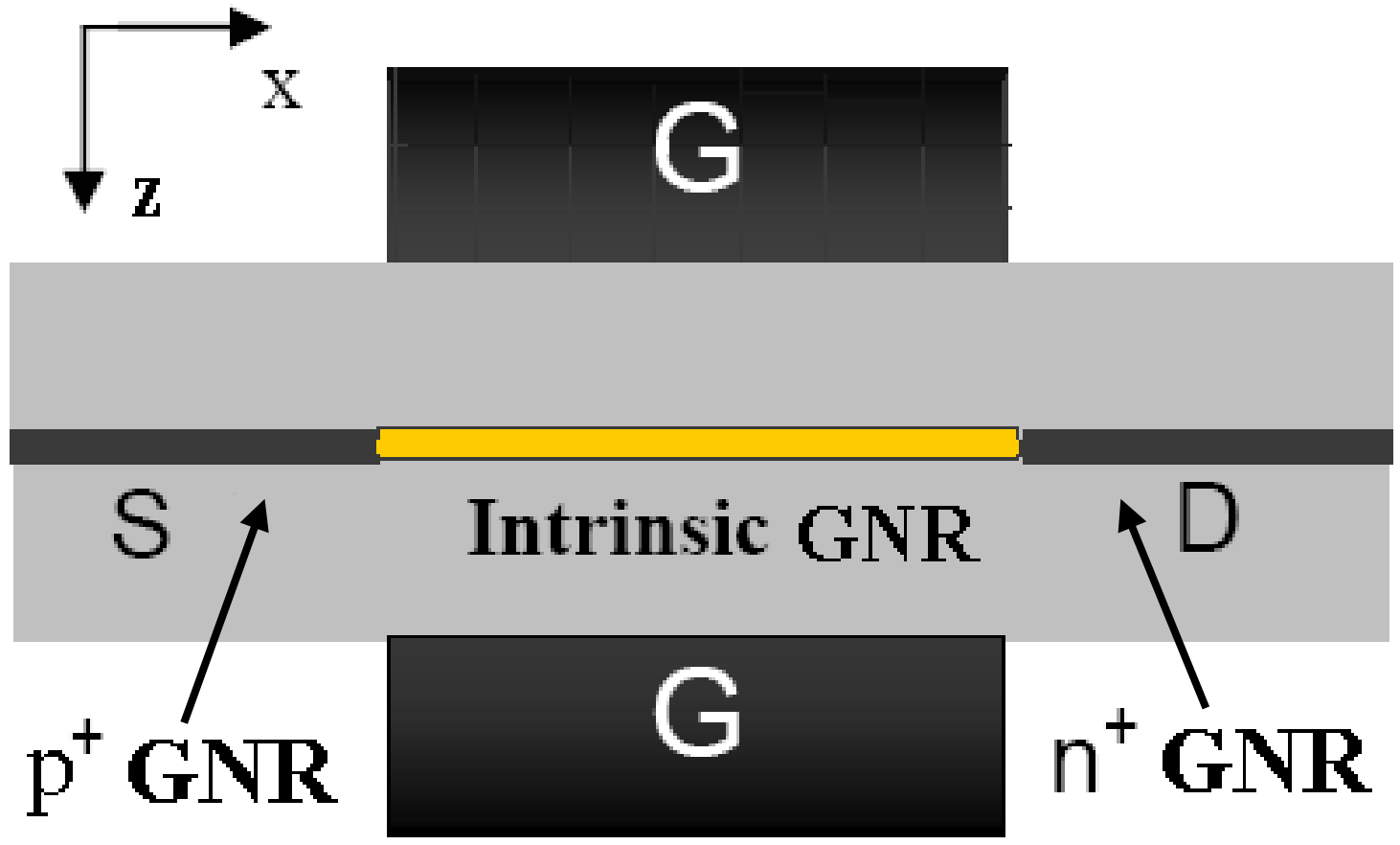

Figure 1

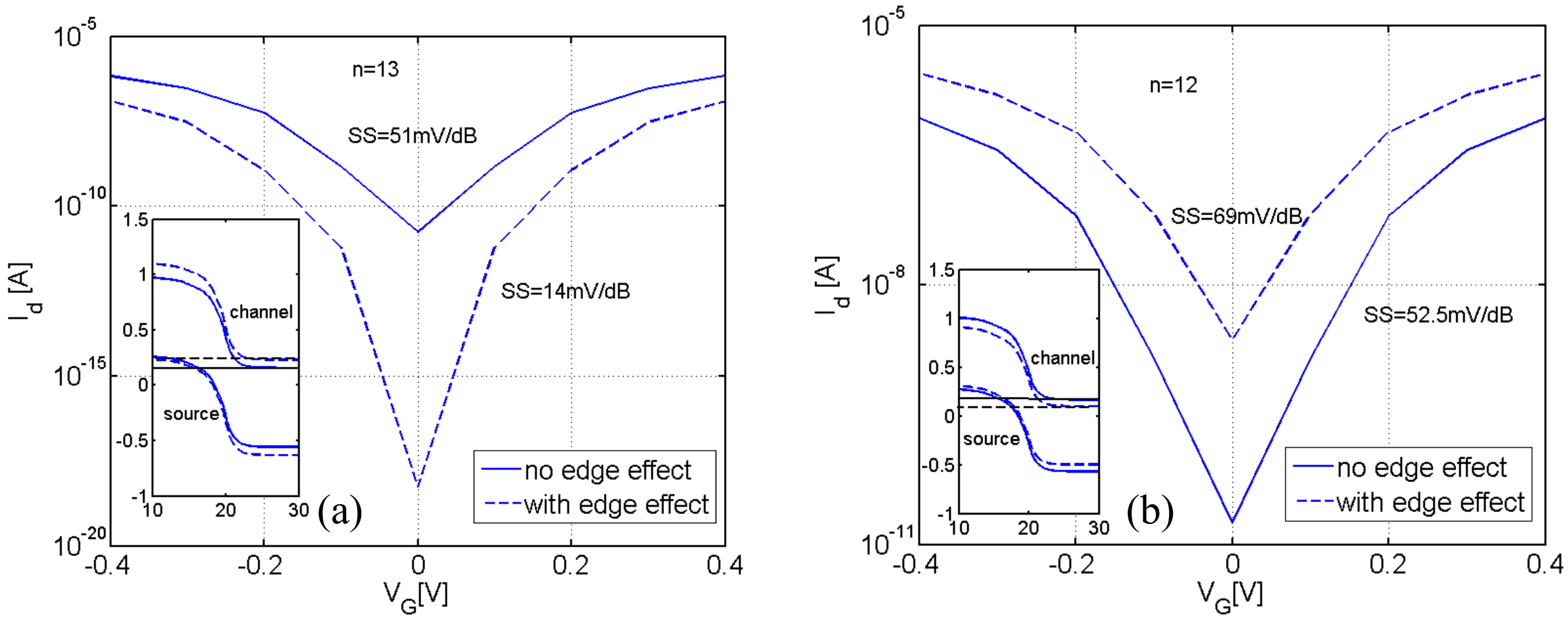
n=13
SS=51mV/dB
SS=14mV/dB
channel
source
no edge effect
with edge effect
(a)
I d [A]
V G[V]
n=12
SS=69mV/dB
SS=52.5mV/dB
channel
source
no edge effect
with edge effect
(b)
I d [A]
V G[V]

Figure 2

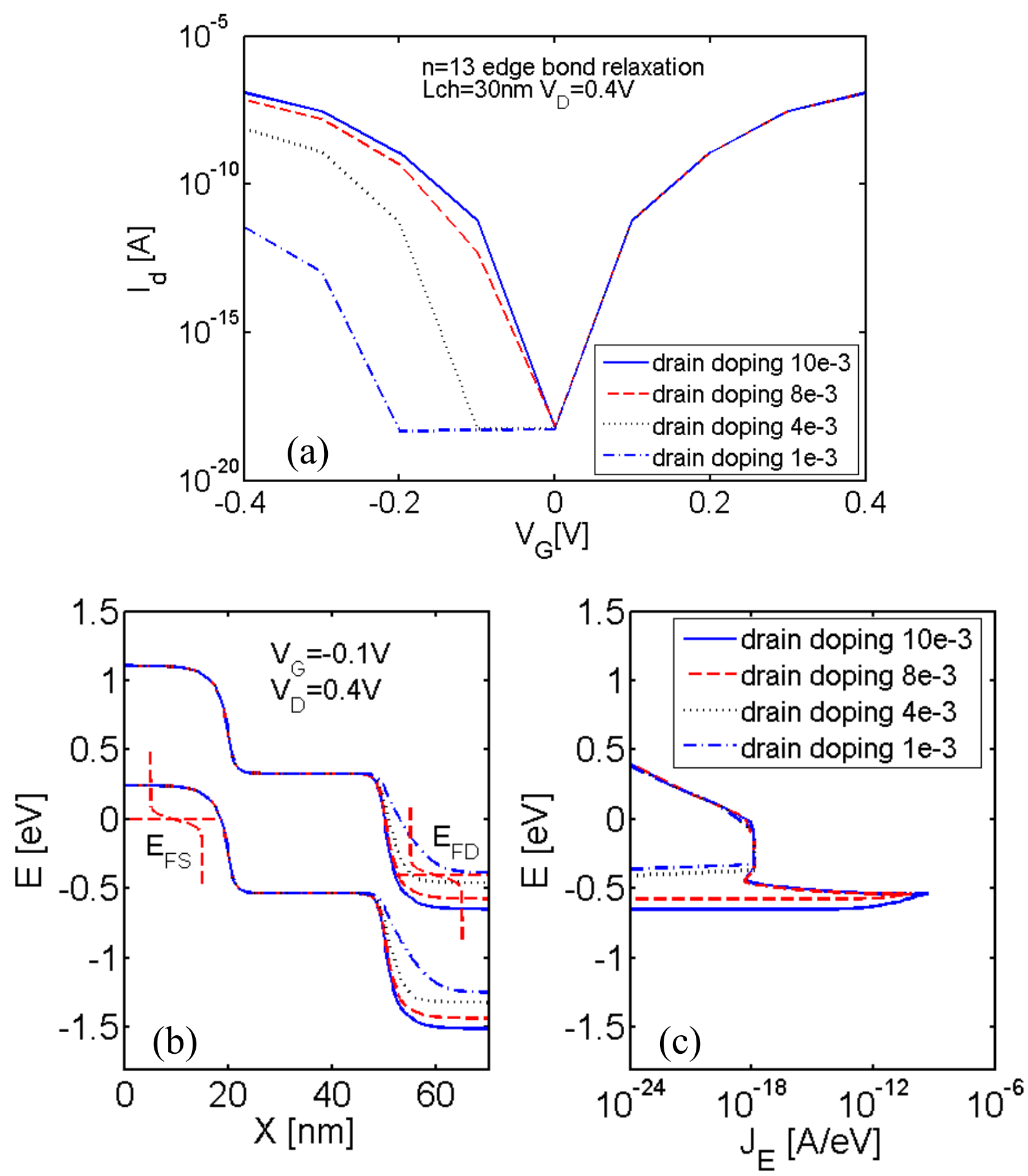
n=13 edge bond relaxation
Lch=30nm $V_D$=0.4V
$I_d$ [A]
$V_G$[V]
drain doping 10e-3
drain doping 8e-3
drain doping 4e-3
drain doping 1e-3
(a)
$V_G$=-0.1V
$V_D$=0.4V
$E_{FS}$
$E_{FD}$
E [eV]
X [nm]
(b)
$J_E$ [A/eV]
(c)

Figure 3

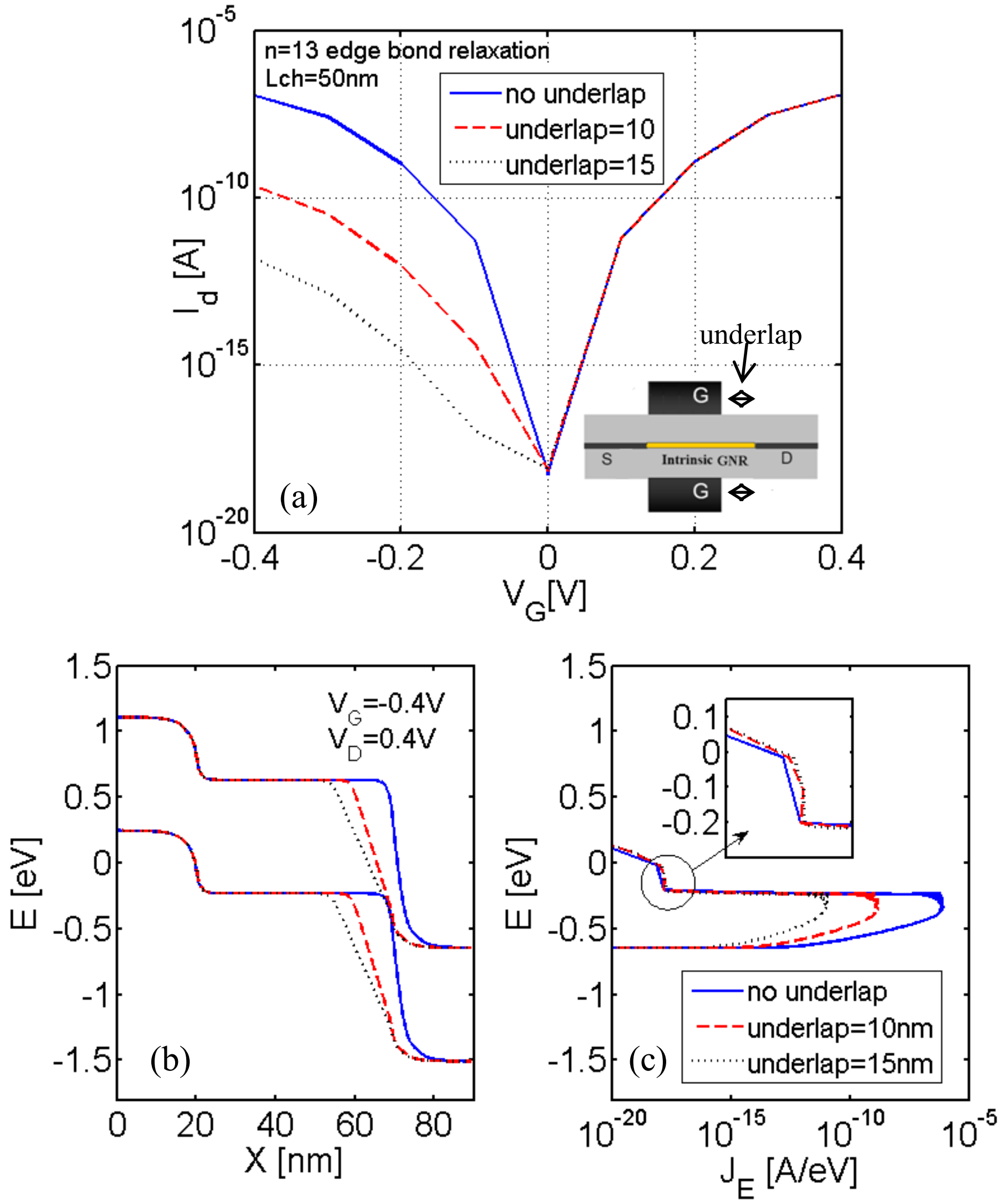
n=13 edge bond relaxation
Lch=50nm
no underlap
underlap=10
underlap=15
underlap
G
S
Intrinsic GNR
D
G
(a)
I_d [A]
V_G[V]
V_G=-0.4V
V_D=0.4V
(b)
E [eV]
X [nm]
no underlap
underlap=10nm
underlap=15nm
(c)
E [eV]
J_E [A/eV]

Figure 4

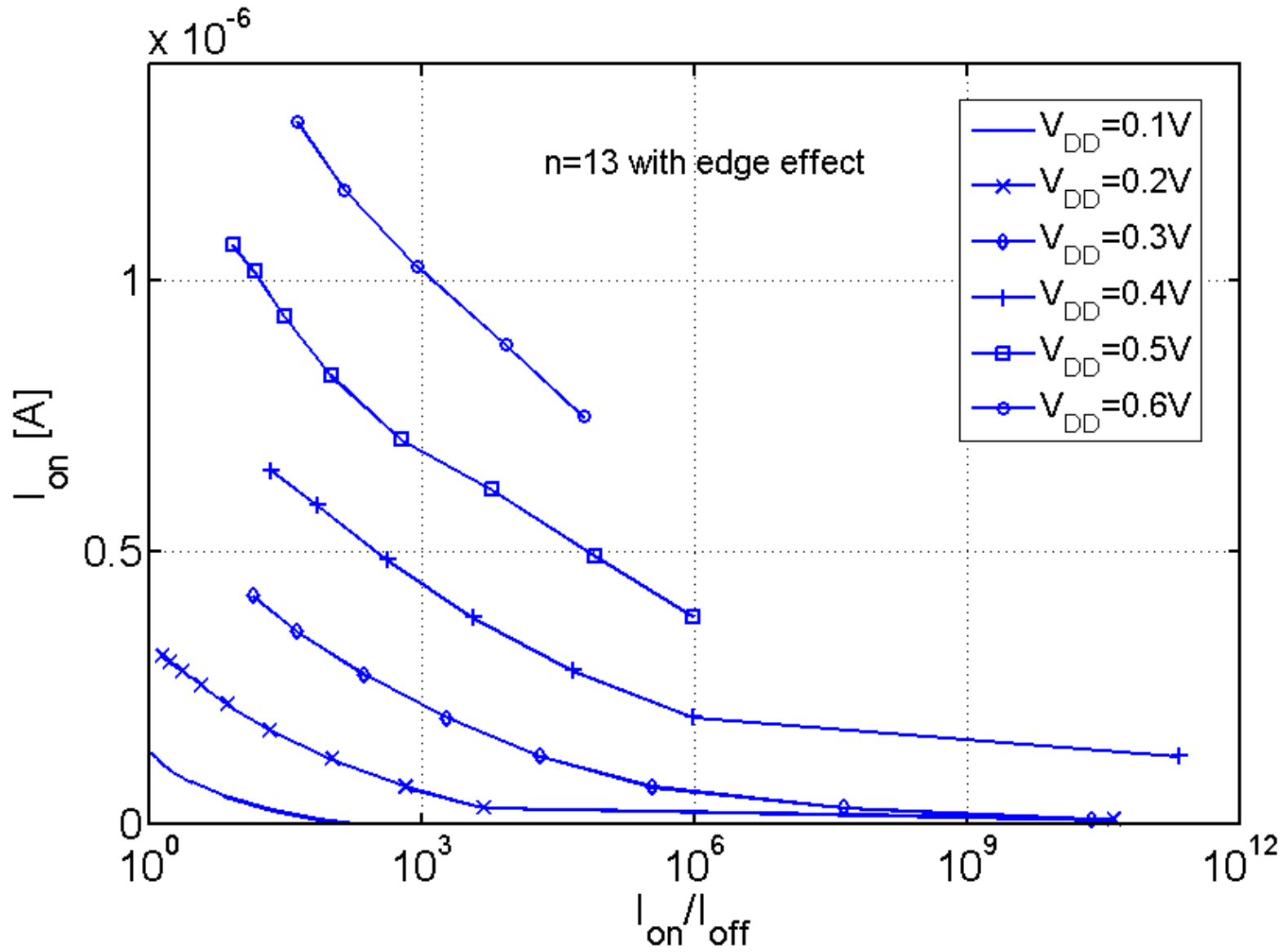
x 10^-6
n=13 with edge effect
V_DD=0.1V
V_DD=0.2V
V_DD=0.3V
V_DD=0.4V
V_DD=0.5V
V_DD=0.6V
I_on [A]
1
0.5
0
10^0
10^3
10^6
10^9
10^12
I_on/I_off

Figure 5